\documentclass[11pt,a4paper]{article}
\usepackage[utf8]{inputenc}
\usepackage{amsmath}
\usepackage{amsfonts}
\usepackage{amssymb}
\usepackage{amsthm}

\numberwithin{equation}{section}
\allowdisplaybreaks

\RequirePackage[charter, greekuppercase=italicized,cal=cmcal,ttscaled=false]{mathdesign}
\usepackage[T1]{fontenc}

\usepackage[colorlinks=true,linkcolor=blue,urlcolor=blue,citecolor=blue,anchorcolor=green,pdfusetitle]{hyperref}
\usepackage{cleveref}

\usepackage[style=alphabetic,doi=false,isbn=false,url=false,maxbibnames=6,maxcitenames=2,giveninits=true,noerroretextools]{biblatex}
\renewbibmacro{in:}{}

\usepackage{fullpage}
\usepackage[font=small,labelfont=bf]{caption}
\newtheorem{thm}{Theorem}\crefname{thm}{Theorem}{Theorems}
\crefname{prop}{Proposition}{Propositions}
\newtheorem{lem}[thm]{Lemma}\crefname{lem}{Lemma}{Lemmas}
\crefname{cor}{Corollary}{Corollaries}

\newtheorem*{support-vector-psd}{\Cref{lem:support-vector-psd}}\crefname{support-vector-psd}{Lemma}{Lemmas}

\usepackage{mathtools}
\usepackage{dsfont}
\usepackage{xcolor}

\newcommand{\cH}{\mathcal{H}}

\newcommand{\cU}{\mathcal{U}}

\newcommand{\bC}{\mathbb{C}}

\DeclareMathOperator{\tr}{tr}
\DeclareMathOperator{\id}{id}
\DeclareMathOperator{\im}{im} 
\DeclareMathOperator{\supp}{supp}

\newcommand{\eps}{\varepsilon}
\newcommand{\one}{\mathds{1}}
\newcommand{\brho}{\bar{\rho}}

\newcommand{\psucc}{p_{\mathrm{succ}}}
\newcommand{\Fstd}{F^{\mathrm{std}}}

\newcommand{\qqquad}{\qquad\qquad}

\title{Optimality of the pretty good measurement\\ for port-based teleportation}

\author{\href{https://orcid.org/0000-0002-1073-9795}{\color{black}Felix Leditzky}}
\date{\small Department of Mathematics \& IQUIST, University of Illinois Urbana-Champaign\\
Institute for Quantum Computing \& Department of Combinatorics and Optimization, University of Waterloo\\
Perimeter Institute for Theoretical Physics\\[.5em]
Email: \href{mailto:leditzky@illinois.edu}{\texttt{leditzky@illinois.edu}}\\[1em]
\normalsize\today
}

\begin{document}
\maketitle

\begin{abstract}
	Port-based teleportation (PBT) is a protocol in which Alice teleports an unknown quantum state to Bob using measurements on a shared entangled multipartite state called the port state and forward classical communication.
	In this paper, we give an explicit proof that the so-called pretty good measurement, or square-root measurement, is optimal for the PBT protocol with independent copies of maximally entangled states as the port state.
	We then show that the very same measurement remains optimal even when the port state is optimized to yield the best possible PBT protocol.
	Hence, there is one particular pretty good measurement achieving the optimal performance in both cases.
	The following well-known facts are key ingredients in the proofs of these results:
	(i) the natural symmetries of PBT, leading to a description in terms of representation-theoretic data;
	(ii) the operational equivalence of PBT with certain state discrimination problems, which allows us to employ duality of the associated semidefinite programs.
	Along the way, we re\-derive the representation-theoretic formulas for the performance of PBT protocols proved in [Studziński et al., 2017] and [Mozrzymas et al., 2018] using only standard techniques from the representation theory of the unitary and symmetric groups.
	Providing a simplified derivation of these beautiful formulas is one of the main goals of this paper.
\end{abstract}

\section{Introduction}

Quantum teleportation \cite{bennett1993teleportation} is arguably one of the most fundamental quantum information-processing tasks.
Its basic setup consists of two spatially separated parties Alice and Bob with access to the following two resources: a classical communication link, and a shared entangled quantum state.
The goal of teleportation is to use these two resources to teleport an unknown quantum state from Alice to Bob.
In the original protocol by \textcite{bennett1993teleportation}, Alice measures the unknown quantum state together with her half of the shared entangled state, and sends the classical outcome to Bob through the classical communication link.
Bob then applies a suitable correction operation to his half of the shared entangled state, thereby transforming it into the desired target state that is now in his possession.
Provided that both the shared entanglement and the classical communication link are noiseless, the quantum teleportation protocol of \cite{bennett1993teleportation} is perfect: it always works, and it faithfully teleports the unknown quantum state from Alice to Bob.

In certain (e.g., cryptographic) applications, one may be interested in simplifying the correction step of the teleportation protocol described above, at the expense of other features or resources.
We focus here on a variant of teleportation called \emph{port-based teleportation} (PBT), which was introduced by \textcite{ishizaka2008asymptotic,ishizaka2009quantum} as a modification of a linear optics teleportation scheme by \textcite{knill2001scheme}.
In PBT, Alice and Bob share a multipartite entangled state on a collection of quantum systems called ports that are distributed evenly between them.
To teleport an unknown quantum state, Alice again performs a joint measurement on the quantum systems in her possession, consisting of the quantum state to be teleported and her half of the ports.
Alice's measurement results in the teleportation of the target state into one of Bob's ports, which is identified by the measurement outcome.
Once Alice communicates the location of the correct port to Bob, he simply discards the other systems.
A general PBT protocol is completely determined by the multipartite entangled state on the ports as well as Alice's measurement.

The crucial property of a PBT protocol as described above is that it works equally well if Bob applies the same unitary operation to each of his port systems before the protocol starts.
As a result, PBT allows for the teleportation of an unknown quantum state processed by a unitary operation, a property called \emph{unitary covariance}.
Unfortunately, in the case of finite resources such unitarily covariant protocols cannot be perfect \cite{nielsen1997programmable}, and hence PBT can only achieve approximate teleportation.
Nevertheless, there are PBT protocols that become faithful in the limit of a large number of port systems \cite{ishizaka2008asymptotic,ishizaka2009quantum,beigi2011simplified,mozrzymas2018optimal,christandl2021asymptotic}.
The unitary covariance property of PBT enables interesting applications for universal programmable quantum processors \cite{ishizaka2008asymptotic}, instantaneous non-local quantum computation \cite{beigi2011simplified}, linking quantum communication complexity and non-locality \cite{buhrman2016quantum}, quantum channel discrimination \cite{pirandola2019fundamental,pereira2021bounds}, channel simulation \cite{pereira2021characterising}, and high-energy physics \cite{may2019quantum,dolev2022non,may2022complexity}.
Furthermore, PBT has been generalized to a ``multi-port'' version where multiple systems are teleported at once \cite{studzinski2020efficient,kopszak2021multiport,mozrzymas2021optimal}.
The resource requirements of PBT have been further investigated in \cite{studzinski2021degradation,strelchuk2021minimal}.

Port-based teleportation enjoys an equivalent description in terms of a certain state discrimination problem (\cite{ishizaka2008asymptotic,ishizaka2009quantum,beigi2011simplified}; see \Cref{sec:pbt} for details).
This useful equivalence enables the study of PBT using semidefinite programming \cite[Sec.~1.2.3]{watrous2018quantum}, and it furthermore suggests the use of a special measurement called the \emph{pretty good measurement} or \emph{square-root measurement} (\cite{belavkin1975optimal,holevo1979asymptotically,hausladen1994pretty}; see \Cref{sec:pgm} for the definition).
In a generic state discrimination problem, this measurement always achieves a success probability no worse than the square of the optimal success probability \cite{barnum2002reversing}.
In this paper, we will employ the connection to state discrimination and semidefinite programming to show that the pretty good measurement is in fact \emph{optimal} for certain PBT protocols of interest.

\subsection{Main results, purpose, and structure of this paper}

The main result of this paper is an explicit proof of optimality of the pretty good measurement for the port-based teleportation (PBT) protocol using $N$ maximally entangled states.
In addition, we show that, somewhat surprisingly, the \emph{same} pretty good measurement used in the previous result also achieves the optimal entanglement fidelity for a PBT protocol with an optimized port state.
Both results are derived by exploiting the natural symmetries of PBT and using its operational equivalence to state discrimination.
The former leads to the known representation-theoretic formulas for the performance of PBT protocols in the two settings based on $N$ maximally entangled states and an optimized port state, derived by \textcite{studzinski2017port} and \textcite{mozrzymas2018optimal}, respectively.
The equivalence of PBT to state discrimination along with the latter's semidefinite programming formulation then allows us to prove that in both cases above the pretty good measurement is in fact the optimal measurement.

Optimality of the pretty good measurement for $N$ maximally entangled states is implied by the results in \cite{mozrzymas2018optimal}. 
These results can furthermore be used to show that the optimal measurement in the case of an optimized port state has the form of a pretty good measurement \cite{studzinski2020personal}.
The present paper provides \emph{explicit} proofs of both results.
Along the way, based on the insights of our prior work \cite{christandl2021asymptotic} we also present an (almost) self-contained derivation of the beautiful formulas of \cite{studzinski2017port,mozrzymas2018optimal} mentioned above.
One of this paper's main goals is a streamlined presentation of these results that is intended to be accessible to a wide audience.
Our approach is similar in spirit to the original proof method based on so-called partially transposed permutation operators \cite{mozrzymas2018simplified} employed in \cite{studzinski2017port,mozrzymas2018optimal}; 
however, here we only use well-known results about the representation theory of the symmetric and unitary groups such as Schur-Weyl duality (see \Cref{sec:rep-theory}), as well as the results from \cite{christandl2021asymptotic}.

This paper is structured as follows.
In \Cref{sec:preliminaries} we introduce some notation and basic definitions, and we review the necessary facts about Schur-Weyl duality.
\Cref{sec:pbt} introduces PBT and explains the operational equivalence to a certain state discrimination problem.
We then prove our main results: optimality of the pretty good measurement for the PBT protocol using $N$ maximally entangled states in \Cref{sec:standard}, and optimality of the same measurement for the protocol using an optimized port state in \Cref{sec:optimized}.
We conclude in \Cref{sec:discussion} with a discussion of our results and open questions.

\section{Preliminaries}\label{sec:preliminaries}
\subsection{Notation \& definitions}

Quantum systems are associated with finite-dimensional Hilbert spaces $\cH_{A_1}$ labeled by capital letters $A_1$, etc.
A multipartite system $AB$ is associated with the tensor product $\cH_A\otimes \cH_B$.
Given $N$ quantum systems $A^N\equiv A_1\dots A_N$, we use the shortcut $A_i^c \equiv A_1\dots A_{i-1} A_{i+1}\dots A_N$.

A quantum state $\rho_A$ on a quantum system $A$ is a linear positive semidefinite operator on $\cH_A$ with unit trace, $\tr\rho_A = 1$.
A pure state $\psi_A$ on a quantum system $A$ is a state of rank 1, which can be identified with a normalized vector $|\psi\rangle_A\in\cH_A$ such that $\psi_A = |\psi_A\rangle\langle \psi_A|$.
Given a $d$-dimensional quantum system $A$, we use the symbol $\pi_A = \frac{1}{d} \one_A$ for the completely mixed state, where $\one_A$ denotes the identity operator on $\cH_A$.
For a given orthonormal basis $\lbrace |i\rangle_A\rbrace_{i=1}^d$ of a $d$-dimensional quantum system $A$ and an isomorphic system $A'\cong A$, the maximally entangled state $|\Phi^+\rangle_{AA'}$ is defined as
\begin{align}
|\Phi^+\rangle_{AA'} = \frac{1}{\sqrt{d}} \sum_{i=1}^d |i\rangle_A \otimes |i\rangle_{A'},
\end{align}
and satisfies $\tr_{A'} \Phi^+_{AA'} = \pi_A$ (and similarly for $\tr_A \Phi^+_{AA'}$).

A quantum measurement of a quantum system $A$ is described by a positive operator-valued measure (POVM) $E=\lbrace E^i_A\rbrace_{i=1}^N$, which consists of positive semidefinite operators $E^i_A$ satisfying $\sum_{i=1}^N E^i_A = \one_A$.
When measuring the quantum system $A$ in the state $\rho_A$ with the POVM $E$, the outcome $i$ is obtained with probability $\tr(\rho_A E^i_A)$.

We denote by $[X,Y] = XY-YX$ the commutator of two operators $X$ and $Y$.
We will often omit identity operators in expressions involving multiple quantum systems, e.g., $X_{AB} Y_A \equiv X_{AB} (Y_A\otimes \one_B)$, whenever this does not cause confusion.
A partition of $N\in\mathbb{N}$ into $d$ parts is a vector $\mu = (\mu_1,\dots,\mu_d)$ with $\mu_1\geq \dots\geq \mu_d\geq 0$ and $\sum_{i=1}^d \mu_i = N$, and denoted by $\mu\vdash_d N$.
Alternatively, $\mu\vdash_d N$ can be interpreted as a Young diagram whose $i$-th row has $\mu_i$ boxes.
For a given Young diagram $\alpha\vdash_d N-1$, we denote by $\alpha + \square$ a Young diagram obtained by adding a single box to $\alpha$ such that the result is still a Young diagram, i.e., a box may be added to the $i$-th row of $\alpha$ if $\alpha_i < \alpha_{i-1}$.
We denote by $S_N$ the symmetric group of degree $N$, and by $\cU_d$ the group of unitary operators acting on a $d$-dimensional Hilbert space.
For a positive semidefinite operator $X$ with spectral decomposition $X = \sum_{i} \lambda_i |\psi_i\rangle\langle\psi_i|$, the generalized inverse $X^{-1}$ is defined as $X^{-1} \coloneqq \sum_{i\colon \lambda_i > 0} \lambda_i^{-1} |\psi_i\rangle\langle \psi_i|$.
With this definition, $X X^{-1} = X^{-1} X = \Pi_X$, where $\Pi_X \coloneqq \sum_{i\colon \lambda_i > 0} |\psi_i\rangle\langle \psi_i|$ denotes the orthogonal projection onto the support $\supp(X) \coloneqq (\ker X)^\perp$ of $X$.

In our optimality proofs we will make use of the following fact, a version of which appeared in \cite{lewenstein1998separability} (see also \cite{ishizaka2008asymptotic}). We give a proof in \Cref{sec:app-lemma} for the convenience of the reader.

\newcommand{\restateSupportVectorPSD}{
Let $X$ be a positive semidefinite operator on a Hilbert space $\cH$.
For some $K\in\mathbb{N}$ and $c\in\mathbb{R}$ let $\lbrace |\xi_k\rangle \rbrace_{k=1}^K\subset \im(X)$ be a collection of non-zero vectors such that $\langle \xi_j|X^{-1}|\xi_k\rangle = \delta_{j,k} c$ for $1\leq j, k\leq K$. Then, $c>0$ and
\begin{align}
	X \geq \frac{1}{c} \sum_{k=1}^K |\xi_k\rangle\langle \xi_k|.
\end{align}
}

\begin{lem}\label{lem:support-vector-psd}
	\restateSupportVectorPSD
\end{lem}

\subsection{Representation theory of the symmetric and unitary groups}\label{sec:rep-theory}
We consider the representations of $S_N$ and $\cU_d$ on $(\mathbb{C}^d)^{\otimes N}$ by permuting tensor factors and acting diagonally, respectively.
More precisely, the representations are defined by the linear extension of the following actions on product states $|\psi_i\rangle\in \mathbb{C}^d$:
\begin{align}
S_N \ni \pi\colon \bigotimes\nolimits_{i=1}^N |\psi_i\rangle &\longmapsto \bigotimes\nolimits_{i=1}^N |\psi_{\pi^{-1}(i)}\rangle \label{eq:S_n-tensor-action}\\
\cU_d \ni U\colon \bigotimes\nolimits_{i=1}^N |\psi_i\rangle &\longmapsto \bigotimes\nolimits_{i=1}^N U|\psi_i\rangle.\label{eq:U_d-tensor-action}
\end{align}
It is easy to check that these two actions commute, i.e., $\pi U^{\otimes N} |\phi\rangle = U^{\otimes N}  \pi |\phi\rangle$ for all $\pi\in S_N$, $U\in\cU_d$, and $|\phi\rangle \in (\mathbb{C}^d)^{\otimes N}$.
Furthermore, \emph{Schur-Weyl duality} states that these representations span each other's commutant (\cite{simon1996representations,fulton1997young}; see also the PhD theses of \textcite{harrow2005phd} and \textcite{christandl2006phd}).
This fact gives rise to a useful decomposition of $(\bC^d)^{\otimes N}$ when considering the actions of the representations of $S_N$ in \eqref{eq:S_n-tensor-action} and $\cU_d$ in \eqref{eq:U_d-tensor-action} together:
\begin{align}
(\bC^d)^{\otimes N} = \bigoplus_{\mu\vdash_d N} V_\mu^d \otimes W_\mu.
\label{eq:schur-weyl-duality}
\end{align}
Here, for a given Young diagram $\mu\vdash_d N$ each direct summand is the tensor product of the \emph{Weyl module} $V_\mu^d$ carrying an irreducible representation of the unitary group $\cU_d$ labeled by $\mu$, and the \emph{Specht module} $W_\mu$ carrying an irreducible representation of the symmetric group $S_N$, again labeled by $\mu$.
Note that \eqref{eq:schur-weyl-duality} only includes all irreducible representations of $S_N$ if $N\leq d$.
We denote the dimensions of the Weyl and Specht modules by $m_{d,\mu} = \dim V_\mu^d$ and $d_\mu = \dim W_\mu$, respectively.
Throughout the paper, $P_\mu$ denotes the projection onto the direct summand $V_\mu^d\otimes W_\mu$ in \eqref{eq:schur-weyl-duality}.
The following result will be useful for us:

\begin{lem}[Partial trace of Young projectors, {\cite{Audenaert2006,christandl2007one}}] \label{lem:Young-proj-partial-trace}
	Let $\mu \vdash_d N$ be a Young diagram with $N$ boxes and at most $d$ rows, and let $P_\mu$ be the corresponding isotypical projection.
	Then,
	\begin{align}
	\tr_1 P_\mu = m_{d,\mu} \sum_{i\colon \mu_i > \mu_{i+1}} \frac{1}{m_{d,\,\mu-\eps_i}} P_{\mu-\eps_i},
	\end{align}
	where $\tr_1$ denotes the partial trace over the first factor in $(\bC^d)^{\otimes N}$, and $\eps_i$ is the vector of length $d$ with a $1$ in the $i$-th component and zeros elsewhere.
\end{lem}

\section{Port-based teleportation}\label{sec:pbt}

In a general PBT protocol \cite{ishizaka2008asymptotic,ishizaka2009quantum}, Alice and Bob share an entangled state $\phi_{A^NB^N}$ defined on Alice's port systems $A^N$ and Bob's ports $B^N$, where $A_i$ and $B_i$ for $i=1,\dots,N$ are $d$-dimensional quantum systems.
Alice holds an additional $d$-dimensional quantum system $A_0$ that she wishes to teleport to Bob.
To achieve this task, she chooses a POVM $E=\lbrace E^i_{A_0A^N}\rbrace_{i=1}^N$ to measure the systems $A_0A^N$, and communicates the outcome $1\leq i\leq N$ to Bob.
Upon receiving this message, Bob discards all but the $i$-th port, which should now hold an approximate copy $B_0$ of Alice's initial system $A_0$.\footnote{\label{fn:probabilistic}In this paper, we are only concerned with so-called \emph{deterministic} PBT as described above.
There is another variant of the protocol called \emph{probabilistic} PBT, in which the protocol teleports the target state perfectly, but may abort with a certain probability.
Both variants were introduced in the original papers \cite{ishizaka2008asymptotic,ishizaka2009quantum}, and we refer to \cite{christandl2021asymptotic} for a more detailed comparison of the two variants.}
Bob's part of the protocol is equivalent to applying the ``correction operation'' $\tr_{B_i^c}$ to his ports.
This operation commutes with any local unitary $U^{\otimes N}$ for $U\in\cU_d$, and thus leads to the unitary covariance property of PBT mentioned in the introduction \cite{ishizaka2008asymptotic,ishizaka2009quantum,majenz2018phd,christandl2021asymptotic}.

A PBT protocol $(\phi_{A^NB^N},E)$ as introduced above can be described via a teleportation channel $\Lambda\colon A_0\to B_0$.
The output state of the above protocol is given by
\begin{align}
\Lambda(\sigma_{A_0}) = \sum_{i=1}^N \tr_{A_0A^NB_i^c}\left[E^i_{A_0A^N} \left(\sigma_{A_0}\otimes \phi_{A^NB^N}\right)\right],
\label{eq:teleportation-channel}
\end{align}
where in each summand the final port $B_i$ is relabeled as $B_0$.
The quality of a PBT protocol is determined by how close the teleportation channel $\Lambda$ is to the identity channel $\id\colon A_0 \to B_0$.
We quantify this by means of the entanglement fidelity, which measures how well $\Lambda$ preserves correlations with an inaccessible reference system $R\cong A_0\cong B_0$.
The entanglement fidelity is defined as
\begin{align}
F(\Lambda) = \tr\left[\Phi^+_{B_0R} (\Lambda\otimes \id_R) (\Phi^+_{A_0R})\right],
\label{eq:entanglement-fidelity}
\end{align}
and we have $F(\Lambda) = 1$ if and only if $\Lambda$ is the identity channel.
In general, the entanglement fidelity represents an \emph{average} error criterion, whereas the \emph{worst-case} error is quantified by the so-called diamond norm distance on the set of quantum channels.
However, the unitary covariance of PBT \cite{ishizaka2008asymptotic,majenz2018phd,christandl2021asymptotic} renders the two error criteria equivalent \cite{pirandola2019fundamental}, and hence the entanglement fidelity \eqref{eq:entanglement-fidelity} quantifies the worst-case error as well.

\textcite{ishizaka2008asymptotic} (see also \cite{beigi2011simplified}) showed that the entanglement fidelity in \eqref{eq:entanglement-fidelity} can be written as
\begin{align}
F(\Lambda) = \frac{1}{d^2} \sum_{i=1}^N \tr\left( E^i_{A^NB} \sigma^i_{A^NB} \right),
\label{eq:entanglement-fidelity-rewrite}
\end{align}
where the POVM $E=\lbrace E^i_{A_0A^N}\rbrace_{i=1}^N$ from above is now interpreted as a measurement on $A^NB$ with $B\equiv B_0\cong A_0$.
The states $\sigma^i_{A^NB}$ are obtained from the port state $\phi_{A^NB^N}$ as
\begin{align}
\sigma^i_{A^NB} = \tr_{B_i^c} \phi_{A^NB^N}.
\label{eq:sigma-i}
\end{align}

\Cref{eq:entanglement-fidelity-rewrite} shows that the entanglement fidelity $F(\Lambda)$ is in fact proportional to the success probability of distinguishing the states $\sigma^i_{A^NB}$ for $i=1,\dots,N$ drawn uniformly at random.
The (general) state discrimination problem of distinguishing $N$ states $\rho_i$ drawn with (not necessarily uniform) probability $p_i$ for $i=1,\dots,N$ admits the following semidefinite program formulation:
\begin{align}
\psucc = \max \left \lbrace \sum\nolimits_{i=1}^N p_i \tr(\rho_i E_i)\colon E_i\geq 0 \text{ for $i=1,\dots,N$, } \sum\nolimits_{i=1}^N E_i = \one\right\rbrace.
\label{eq:state-discrimination-primal}
\end{align}
We refer to \cite[Sec.~1.2.3]{watrous2018quantum} for an introduction to semidefinite programs.
The dual program of \eqref{eq:state-discrimination-primal} can be derived using standard methods, and is given by the following minimization problem:
\begin{align}
\psucc^* = \min \left\lbrace \tr K\colon K\geq p_i \rho_i \text{ for $i=1,\dots,N$}\right\rbrace.
\label{eq:state-discrimination-dual}
\end{align}
It has the same value as the primal problem \eqref{eq:state-discrimination-primal} by strong duality, $\psucc = \psucc^*$, which follows for example from Slater's Theorem \cite[Sec.~1.2.3]{watrous2018quantum}.

\begin{figure}
	\centering
	\includegraphics{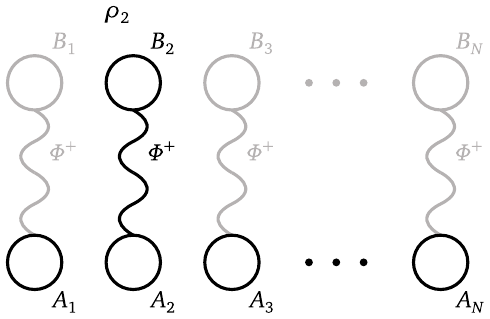}
	\caption{Schematic depiction of the state discrimination problem that is equivalent to PBT as explained in \Cref{sec:pbt}.
		Shown here is the state $\rho_2$ from the family $\lbrace \rho_{i}\rbrace_{i=1}^N$ defined in \eqref{eq:rho_i} that appears in a PBT protocol using $N$ maximally entangled states $\Phi^+_{AB}$.
		The latter are represented by wavy lines.}
	\label{fig:rho-i}
\end{figure}

Using \eqref{eq:entanglement-fidelity-rewrite} and \eqref{eq:state-discrimination-primal}, it is now clear that the entanglement fidelity of a PBT protocol with teleportation channel $\Lambda$ can be expressed as \cite{ishizaka2008asymptotic,beigi2011simplified}
\begin{align}
F(\Lambda) = \frac{N}{d^2} \psucc,
\label{eq:PBT-state-discrimination}
\end{align}
where $\psucc$ is defined in terms of the $N$ states $\sigma^i_{A^NB}$ in \eqref{eq:sigma-i} drawn uniformly at random.
\Cref{fig:rho-i} shows a schematic description of these states when the port state $\phi_{A^NB^N}$ is comprised of $N$ maximally entangled states, as discussed in \Cref{sec:standard}.
\Cref{eq:PBT-state-discrimination} forges a useful operational equivalence between PBT and state discrimination.
We will make use of this equivalence, in particular the semidefinite programming formulation and duality, to derive our main results.
Throughout the discussion, the local port dimension $d$ and the number of ports $N$ are fixed but arbitrary.

\section{Independent maximally entangled states}\label{sec:standard}
We first consider a special case of PBT where the port state is comprised of $N$ independent maximally entangled states,
\begin{align}
\phi_{A^NB^N} = \left(\Phi^+_{AB}\right)^{\otimes N}.
\end{align}
According to \Cref{sec:pbt}, we can equivalently consider the state discrimination problem of distinguishing the $N$ states
\begin{align}
\rho_i = \Phi^+_{A_iB} \otimes \pi_{A_i^c} \label{eq:rho_i}
\end{align}
on $A^NB$ drawn uniformly at random, i.e., with probability $\frac{1}{N}$ each.
A graphical representation of these states is shown in \Cref{fig:rho-i}.

Since $(U\otimes \bar{U}) |\Phi^+\rangle = |\Phi^+\rangle$ for every unitary $U\in\cU_d$,\footnote{Here, $\bar{X}$ denotes complex conjugation with respect to the basis used to define $|\Phi^+\rangle$.} the states $\rho_i$ have the symmetries
\begin{align}
\left[U^{\otimes N} \otimes \bar{U},\rho_i \right] &= 0 \quad\text{for all $U\in \cU_d$,} \label{eq:rho_i-symmetry-U}\\
\left[\varphi\otimes \one_{A_i B},\rho_i\right] &= 0 \quad\text{for all $\varphi\in S_{N-1}$,} \label{eq:rho_i-symmetry-SN-1}
\intertext{where in the first line $U^{\otimes N} \otimes \bar{U} \equiv U^{\otimes N}_A \otimes \bar{U}_B$, and in the second line we consider the action of $S_{N-1}$ on $A_i^c$ by permuting tensor factors.
	Moreover,}
\pi \rho_i \pi^\dagger &= \rho_{\pi(i)} \quad\text{for all $\pi\in S_N$.} \label{eq:rho_i-orbit}
\end{align}
It follows from \cref{eq:rho_i-symmetry-U,eq:rho_i-symmetry-SN-1,eq:rho_i-orbit} that the (unnormalized) average state $\bar{\rho} = \sum_{i=1}^N \rho_i$ on $A^NB$ has the symmetries
\begin{align}
\left[U^{\otimes N} \otimes \bar{U},\brho \right] &= 0 \quad\text{for all $U\in \cU_d$,} \label{eq:brho-symmetry-U}\\
\left[\pi\otimes \one_B,\brho\right] &= 0 \quad\text{for all $\pi\in S_{N}$.} \label{eq:brho-symmetry-SN}
\end{align}

The symmetries in \cref{eq:brho-symmetry-SN,eq:brho-symmetry-U} together with Schur's Lemma imply that $\brho$ is diagonal with respect to the following decomposition of $(\bC^d)^{\otimes N+1}$ derived from Schur-Weyl duality \eqref{eq:schur-weyl-duality} using the so-called dual Pieri rule:
\begin{align}
(\bC^d)^{\otimes N+1} = \bigoplus_{\mu\vdash_d N}\, \bigoplus_{i\colon \mu_i > \mu_{i+1}} V_{\mu-\eps_i}^d \otimes W_\mu.
\label{eq:Cd-N+1-decomposition}
\end{align}
As in \Cref{lem:Young-proj-partial-trace}, $\eps_i$ is the vector of length $d$ with a $1$ in the $i$-th component and zeros elsewhere.\footnote{
	\label{fn:no-YD}Note that we set $\mu_{d+1} = -\infty$ in \eqref{eq:Cd-N+1-decomposition}, and hence the summand for $i=d$ always appears in the sum.
	For a Young diagram $\mu\vdash_d N$ with $\mu_d = 0$ the resulting $\mu-\eps_d$ is not a Young diagram anymore.}
We refer to Appendix A of \cite{christandl2021asymptotic} for details of the derivation of \eqref{eq:Cd-N+1-decomposition}.
In the present paper, we will make use of this result in the following way:

\begin{lem}[{\cite{studzinski2017port,christandl2021asymptotic}}] \label{lem:eigenvalues-average-state}
	The (unnormalized) average state $\bar{\rho} = \sum_{i=1}^N \rho_i$ of the ensemble $\lbrace (\frac{1}{N},\rho_i) \rbrace_{i=1}^N$ with $\rho_i$ as in \eqref{eq:rho_i} can be written as
	\begin{align}
	\bar{\rho} = \bigoplus_{\alpha\vdash_d N-1}\, \bigoplus_{\mu = \alpha + \square} r_{\mu,\alpha}\, \one_{V_\alpha^d} \otimes \one_{W_\mu},
	\end{align}
	where the eigenvalues $r_{\mu,\alpha}$ are given by
	\begin{align}
	r_{\mu,\alpha} = \frac{N}{d^N} \frac{m_{d,\mu} d_\alpha}{m_{d,\alpha} d_\mu}.\label{eq:r-eigenvalues}
	\end{align}
\end{lem}

\subsection{Performance of the pretty good measurement}\label{sec:pgm}

For a given state ensemble $\lbrace (p_i,\sigma_i) \rbrace_{i=1}^N$, the \emph{pretty good measurement} \cite{belavkin1975optimal,holevo1979asymptotically,hausladen1994pretty} is defined as the measurement $E = \lbrace E_i\rbrace_{i=1}^N$ with operators
\begin{align}
E_i &= \bar{\sigma}^{-1/2} \, p_i \sigma_i\, \bar{\sigma}^{-1/2}.
\label{eq:pretty-good-measurement}
\end{align}
Here, $\bar{\sigma} = \sum_{i=1}^N p_i \sigma_i$ is the ensemble average state.
The measurement operators $E_i$ satisfy $E_i\geq 0$ for all $i=1,\dots,N$, and $\sum_{i=1}^N E_i = \Pi_{\bar{\sigma}}$.
The pretty good measurement thus forms a valid POVM once the Hilbert space is restricted to $\supp \bar{\sigma}$, which we will always assume.

The success probability of discriminating the states $\lbrace (\frac{1}{N},\rho_i) \rbrace_{i=1}^N$ with $\rho_i$ as in \eqref{eq:rho_i} using the pretty good measurement $E = \lbrace E_i\rbrace_{i=1}^N$ is thus given by the expression
\begin{align}
\psucc = \frac{1}{N} \sum_{i=1}^N \tr\left(\rho_i \brho^{-1/2} \rho_i \brho^{-1/2}\right),
\label{eq:pgm-psucc}
\end{align}
where as before $\brho=\sum_{i=1}^N \rho_i$ is the \emph{unnormalized} ensemble average state.
It follows from the results of \textcite{studzinski2017port} that this success probability can be expressed in terms of representation-theoretic quantities:
\begin{align}
\psucc = \frac{1}{Nd^N} \sum_{\alpha\vdash_d N-1} \left(\sum_{\mu = \alpha + \square} \sqrt{m_\mu d_\mu}\right)^2.
\label{eq:psucc-pgm-rep-theory}
\end{align}
The goal of this section is to rederive this formula.

To this end, we define the operator
\begin{align}
X = \sum_{i=1}^N \rho_i \brho^{-1/2} \rho_i \brho^{-1/2},
\label{eq:X}
\end{align}
such that $\frac{1}{N}\tr X = \psucc$ for the success probability defined in \eqref{eq:pgm-psucc}.
Since $x\mapsto x^{-1/2}$ is a real-analytic function on $(0,\infty)$, the operator $\brho^{-1/2}$ inherits the $U^{\otimes N}\otimes \bar{U}$ and $S_N$ symmetries (\cref{eq:brho-symmetry-U,eq:brho-symmetry-SN}) from $\brho$.
Furthermore, for any $\pi\in S_N$,
\begin{align}
\pi \rho_i \brho^{-1/2} \rho_i \brho^{-1/2} \pi^\dagger &= \pi \rho_i \pi^\dagger \pi \brho^{-1/2} \pi^\dagger \pi \rho_i \pi^\dagger \pi \brho^{-1/2} \pi^\dagger\\
&= \rho_{\pi(i)} \brho^{-1/2} \rho_{\pi(i)} \brho^{-1/2},
\end{align}
where we used \eqref{eq:rho_i-orbit} and \eqref{eq:brho-symmetry-SN}.

The operator $X$ in \eqref{eq:X} thus has the same $U^{\otimes N}\otimes \bar{U}$ and $S_N$ symmetries as $\brho$ above,
\begin{align}
\left[U^{\otimes N} \otimes \bar{U},X \right] &= 0 \quad\text{for all $U\in \cU_d$,} \label{eq:X-symmetry-U}\\
\left[\pi\otimes \one_B,X\right] &= 0 \quad\text{for all $\pi\in S_{N}$.} \label{eq:X-symmetry-SN}
\end{align}
With respect to the decomposition \eqref{eq:Cd-N+1-decomposition}, the operator $X$ can hence be written as
\begin{align}
X = \bigoplus_{\mu\vdash_d N}\, \bigoplus_{i\colon \mu_i > \mu_{i+1}} x_{\mu,i}\, \one_{V_{\mu-\eps_i}^d} \otimes \one_{W_\mu}.
\label{eq:X-decomposition-1}
\end{align}

The coefficients $x_{\mu,i}$ in \eqref{eq:X-decomposition-1} can be determined using a similar strategy as in Appendix~A of \cite{christandl2021asymptotic}.
Since $d$ is fixed throughout the discussion, we abbreviate $m_\mu\equiv m_{d,\mu}$ for the dimension of the Weyl modules $V_{\mu}^d$ in the following.
Recall that $P_\mu$ denotes the projection onto the summand $V_\mu^d\otimes W_\mu$ in the Schur-Weyl decomposition \eqref{eq:schur-weyl-duality}.
We further denote by $Q_\alpha$ the isotypical projections for the $\cU_d$ action by $U^{\otimes N}\otimes \bar{U}$ as defined via decomposition \eqref{eq:Cd-N+1-decomposition}. 
Note that here $\alpha$ can have negative entries and is thus not necessarily a valid Young diagram (see Footnote \ref{fn:no-YD}).

However, the coefficients $x_{\mu,i}$ in \eqref{eq:X-decomposition-1} are only non-zero when $\mu-\eps_i$ is indeed a valid Young diagram, $\mu-\eps_i=\alpha\vdash_d N-1$.
To see this, we recall the following argument from \cite[App.~A]{christandl2021asymptotic}: Since $(U\otimes \bar{U})|\Phi^+\rangle_{A_1B}$ for every $U\in\cU_d$, the actions of $U^{\otimes N}\otimes \bar{U}$ and $\one_{A_1B}\otimes U^{\otimes N-1}$ agree on the range of $\Phi^+_{A_1B}$, and hence,
\begin{align}
\Phi^+_{A_1B} Q_{\mu-\eps_i} = \begin{cases} \Phi^+_{A_1B}(\one_{A_1B} \otimes P'_\alpha) & \text{if $\alpha = \mu-\eps_i$ is a Young diagram,}\\
0 & \text{otherwise.}
\end{cases}
\label{eq:U-action-Phi+}
\end{align}
Here, $P'_\alpha$ denotes the isotypical projection with respect to the action of $\cU_d$ on $A_1^c$ by $U^{\otimes N-1}$.
The operator $X$ is proportional to a sum of terms of the form $\Phi^+_{A_iB} \brho^{-1/2}\Phi^+_{A_iB} \brho^{-1/2}$, so we can apply the above argument to the coefficients $x_{\mu,i}$ appearing in \eqref{eq:X-decomposition-1} to infer that $x_{\mu,i} \neq 0$ only when $\mu-\eps_i=\alpha$ is a Young diagram. 
We denote these coefficients by $x_{\mu,\alpha}$ henceforth, and write $X$ as
\begin{align}
X = \bigoplus_{\alpha\vdash_d N-1} \, \bigoplus_{\mu = \alpha + \square} x_{\mu,\alpha}\, \one_{V_{\alpha}^d} \otimes \one_{W_\mu}.
\label{eq:X-decomposition}
\end{align}
In the remainder of this subsection, we first compute the trace of this operator, and then derive a formula for the coefficients $x_{\mu,\alpha}$.

Let $\alpha\vdash_d N-1$ and $\mu = \alpha + \square$. 
By symmetry and the form of $\rho_1$ in \eqref{eq:rho_i}, we have
\begin{align}
\tr \left[X\, (P_\mu\otimes \one_B) Q_{\alpha}\right] &= N \tr\left[\rho_1 \brho^{-1/2} \rho_1 \brho^{-1/2} (P_\mu\otimes \one_B) Q_{\alpha}\right]\\
&= \frac{N}{d^{2N-2}} \tr\left[ \Phi^+_{A_1B} \brho^{-1/2} \Phi^+_{A_1B} \brho^{-1/2} (P_\mu\otimes \one_B) Q_{\alpha}\right]\\
&= \frac{N}{d^{2N-2}} \sum_{\alpha',\alpha''\vdash_d N-1}\, \sum_{\substack{\mu'=\alpha'+\square\\ \mu''=\alpha''+\square}} r_{\mu',\alpha'}^{-1/2} r_{\mu'',\alpha''}^{-1/2}\notag\\
&\qquad {} \times \tr\left[ \Phi^+_{A_1B} (P_{\mu'} \otimes \one_B) Q_{\alpha'} \Phi^+_{A_1B} (P_{\mu''}\otimes \one_B) Q_{\alpha''} (P_\mu\otimes \one_B) Q_{\alpha} \right],\label{eq:x-coefficients-inter1}
\end{align}
where we inserted the decomposition of $\brho$ from \Cref{lem:eigenvalues-average-state} twice in \eqref{eq:x-coefficients-inter1}.
Let us take a closer look at the trace quantity in \eqref{eq:x-coefficients-inter1} for fixed $\alpha\vdash_d N-1$ and $\mu = \alpha + \square$. 
Using the fact that $P_*\otimes \one_B$ commutes with $Q_*$, we can apply the identity \eqref{eq:U-action-Phi+} to each of $Q_{\alpha}$, $Q_{\alpha'}$ and $Q_{\alpha''}$ to obtain
\begin{align}
&\tr\left[ \Phi^+_{A_1B} (P_{\mu'} \otimes \one_B) Q_{\alpha'} \Phi^+_{A_1B} (P_{\mu''}\otimes \one_B) Q_{\alpha''} (P_\mu\otimes \one_B) Q_{\alpha} \right] \notag\\
&\qqquad {} = \tr\left[ \Phi^+_{A_1B} (P_{\mu'} \otimes \one_B) \Phi^+_{A_1B} (P_{\mu''}P_\mu\otimes \one_B) (\one_{A_1B} \otimes P'_{\alpha'}P'_{\alpha''}P'_{\alpha}) \right]\\
&\qqquad {} = \tr\left[ \Phi^+_{A_1B} (P_{\mu'} \otimes \one_B) \Phi^+_{A_1B} (P_{\mu}\otimes \one_B)(\one_{A_1B} \otimes P'_{\alpha}) \right] \delta_{\mu,\mu''} \delta_{\alpha,\alpha'} \delta_{\alpha,\alpha''}, \label{eq:x-coefficients-inter15}
\end{align}
where we used orthogonality among the projectors $P_{*}$ and among the $P'_*$ in the last line.
Substituting \eqref{eq:x-coefficients-inter15} in \eqref{eq:x-coefficients-inter1} leads to
\begin{align}
\tr \left[X (P_\mu\otimes \one_B) Q_{\alpha}\right] &= \frac{N}{d^{2N-2}} r_{\mu,\alpha}^{-1/2} \sum_{\mu'=\alpha+\square} r_{\mu',\alpha}^{-1/2} \tr\left[ \Phi^+_{A_1B} (P_{\mu'} \otimes \one_B) \Phi^+_{A_1B} (P_\mu\otimes \one_B) (\one_{A_1B}\otimes P'_{\alpha}) \right]\label{eq:x-coefficients-inter3}\\
&= \frac{N}{d^{2N}} r_{\mu,\alpha}^{-1/2} \sum_{\mu'=\alpha+\square} r_{\mu',\alpha}^{-1/2} \tr\left[ \tr_{A_1} (P_{\mu'}) \tr_{A_1}( P_\mu) P'_{\alpha} \right],\label{eq:x-coefficients-inter4}
\end{align} 
where step \eqref{eq:x-coefficients-inter4} follows from the elementary identity $\tr\left[\Phi^+_{RS} X_{ST} \Phi^+_{RS} Y_{ST}\right] = \frac{1}{d^2} \tr \left(X_T Y_T\right)$.

For the partial traces of the Young projectors $P_*$, \Cref{lem:Young-proj-partial-trace} gives 
\begin{align}
\tr_{A_1} (P_\mu) P'_\alpha = \frac{m_\mu}{m_\alpha} P'_{\alpha},
\end{align}
and similarly, $\tr_{A_1} (P_{\mu'}) P'_\alpha = m_{\mu'} m_\alpha^{-1} P'_{\alpha}$.
Substituting these two relations together with formula \eqref{eq:r-eigenvalues} for the coefficients $r_{\mu,\alpha}$ in \eqref{eq:x-coefficients-inter4} yields
\begin{align}
\tr \left[X (P_\mu\otimes \one_B) Q_{\alpha}\right] &= \frac{N}{d^{2N}} \frac{d^N}{N} \frac{m_\alpha}{d_\alpha} \frac{\sqrt{d_\mu}}{\sqrt{m_\mu}} \sum_{\mu'=\alpha + \square} \frac{\sqrt{d_{\mu'}}}{\sqrt{m_{\mu'}}} \frac{ m_{\mu} m_{\mu'}}{m_\alpha^2} \tr P'_\alpha\\
&= \frac{1}{d^N} \sqrt{d_\mu m_{\mu}} \sum_{\mu'=\alpha + \square} \sqrt{d_{\mu'} m_{\mu'}}.\label{eq:x-coefficients-inter5}
\end{align}
Summing over $\alpha\vdash_d N-1$ and $\mu = \alpha + \square$ gives the trace of $X$,
\begin{align}
\tr X &= \sum_{\alpha\vdash_d N-1}\sum_{\mu=\alpha + \square} \tr \left[X (P_\mu\otimes \one_B) Q_{\alpha}\right]\\
&= \frac{1}{d^N} \sum_{\alpha\vdash_d N-1} \left(\sum_{\mu=\alpha + \square} \sqrt{d_{\mu} m_{\mu}}\right)^2,
\end{align}
and thus we have proved \eqref{eq:psucc-pgm-rep-theory} via $\psucc = \frac{1}{N}\tr X$.

It remains to derive a formula for the coefficients $x_{\mu,\alpha}$ appearing in \eqref{eq:X-decomposition}.
By definition, for $\alpha\vdash_d N-1$ and $\mu = \alpha + \square$ we have
\begin{align}
\tr \left[X (P_\mu\otimes \one_B) Q_{\alpha}\right] = x_{\mu,\alpha} m_{\alpha} d_\mu,
\end{align}
which is equal to \eqref{eq:x-coefficients-inter5} by the above calculation.
Thus,
\begin{align}
x_{\mu,\alpha} = \frac{1}{d^N} \frac{\sqrt{m_\mu}}{\sqrt{d_\mu}} \frac{1}{m_\alpha} \sum_{\mu' = \alpha + \square} \sqrt{d_{\mu'}m_{\mu'} }.
\label{eq:x-coefficients-formula}
\end{align}

\subsection{Optimality of the pretty good measurement}\label{sec:standard-optimality}

To prove optimality of the pretty good measurement for distinguishing the states $\rho_i$ defined in \eqref{eq:rho_i}, we use the dual program \eqref{eq:state-discrimination-dual} of the corresponding state discrimination problem.
The proof idea is identical to the method used by \textcite{ishizaka2008asymptotic} to prove optimality of the pretty good measurement for qubit port systems.
Recall from the previous section that $\psucc = \frac{1}{N} \tr X$ for the operator $X$ defined in \eqref{eq:X}.
If we can show that $\frac{1}{N} X$ is feasible for the dual program \eqref{eq:state-discrimination-dual}, then optimality follows from the fact that any feasible solution to \eqref{eq:state-discrimination-dual} is an upper bound on the optimal solution, given by either \eqref{eq:state-discrimination-primal} or \eqref{eq:state-discrimination-dual} because of strong duality.
By construction, this upper bound is identical to the value of the success probability \eqref{eq:psucc-pgm-rep-theory} calculated in \Cref{sec:pgm}, which establishes optimality of the pretty good measurement.

Feasibility of the operator $\frac{1}{N} X$ for \eqref{eq:state-discrimination-dual} is equivalent to showing that $X \geq \rho_i$ for all $i=1,\dots,N$; by symmetry, it is enough to prove this for $\rho_1 = \Phi^+_{A_1B}\otimes \pi_{A_1^c}$.
This will follow from \Cref{lem:support-vector-psd} applied to the operator $X$ and a set of vectors into which $\rho_1$ can be decomposed.
First, we recall the expression $
	X = \bigoplus_{\alpha\vdash_d N-1} \, \bigoplus_{\mu = \alpha + \square} x_{\mu,\alpha}\, \one_{V_{\alpha}^d} \otimes \one_{W_\mu}
$
derived in \Cref{sec:pgm}.
Formula \eqref{eq:x-coefficients-formula} for the coefficients $x_{\mu,\alpha}$ shows that they are strictly positive for all $\alpha\vdash_d N-1$ and $\mu=\alpha + \square$, so that $X$ is manifestly positive definite.
For the set of vectors in the statement of \Cref{lem:support-vector-psd} we choose the following eigenvectors of $\rho_1 = \Phi^+_{A_1B}\otimes \pi_{A_1^c}$,
\begin{align}
	|\xi(\alpha,q_\alpha,p_\alpha)\rangle \coloneqq |\Phi^+\rangle_{A_1B} \otimes |\alpha,q_\alpha,p_\alpha\rangle_{A_1^c}
\end{align}
for $\alpha\vdash_d N-1$, $1\leq q_\alpha\leq m_\alpha$, and $1\leq p_\alpha\leq d_\alpha$.
Here, $\lbrace |\alpha,q_\alpha,p_\alpha\rangle_{A_1^c} \rbrace_{\alpha\vdash_d N-1,q_\alpha,p_\alpha}$ is the \emph{Schur basis} \cite{harrow2005phd,bacon2006efficient} adapted to the Schur-Weyl decomposition $(\bC^d)^{\otimes N-1} = \bigoplus_{\alpha\vdash_d N-1} V_\alpha^d \otimes W_\alpha$.
The indices $1\leq q_\alpha\leq m_\alpha$ and $1\leq p_\alpha\leq d_\alpha$ correspond to the Weyl module $V_\alpha^d$ and the Specht module $W_\alpha$, respectively.
Since
\begin{align}
	\rho_1 = \frac{1}{d^{N-1}} \Phi^+_{A_1B} \otimes \one_{A_1^c} = \frac{1}{d^{N-1}} \sum_{\alpha\vdash_d N-1} \sum_{q_\alpha,p_\alpha} \Phi^+_{A_1B} \otimes |\alpha,q_\alpha,p_\alpha\rangle\langle \alpha,q_\alpha,p_\alpha|_{A_1^c},
	\label{eq:rho-in-schur-basis}
\end{align}
the desired operator inequality $X\geq \rho_1$ follows from \Cref{lem:support-vector-psd} once we establish that
\begin{align}
	\langle\xi(\alpha,q_{\alpha},p_{\alpha}) |\,X^{-1}|\xi(\beta,\tilde{q}_\beta,\tilde{p}_\beta)\rangle = \delta_{\alpha,\beta} \delta_{q_\alpha,\tilde{q}_\alpha} \delta_{p_\alpha,\tilde{p}_\alpha} d^{N-1}\label{eq:inverse-innerproduct}
\end{align}
holds for all $\alpha,\beta\vdash_d N-1$, $1\leq q_\alpha\leq m_\alpha$, $1\leq \tilde{q}_\beta \leq m_\beta$, $1\leq p_\alpha\leq d_\alpha$, and $1\leq \tilde{p}_\beta\leq d_\beta$.

To this end, we compute:
\begin{align}
&\langle\xi(\alpha,q_{\alpha},p_{\alpha}) |\,X^{-1}|\xi(\beta,\tilde{q}_\beta,\tilde{p}_\beta)\rangle\notag\\
&\qquad {} = \sum_{\alpha'\vdash_d N-1}\, \sum_{\mu = \alpha'+\square} x_{\mu,\alpha'}^{-1} \tr\left[ \left( P_{\mu}\otimes \one_B \right) Q_{\alpha'} \left(\Phi^+_{A_1B} \otimes |\beta,\tilde{q}_\beta,\tilde{p}_\beta\rangle \langle \alpha,q_\alpha,p_\alpha|_{A_1^c}\right)\right]\\
&\qquad {} = \sum_{\alpha'\vdash_d N-1}\, \sum_{\mu = \alpha'+\square} x_{\mu,\alpha'}^{-1} \tr\left[ \left( P_{\mu}\otimes \one_B \right) \left(\Phi^+_{A_1B} \otimes P'_{\alpha'}|\beta,\tilde{q}_\beta,\tilde{p}_\beta\rangle \langle \alpha,q_\alpha,p_\alpha|_{A_1^c}\right)\right] \label{eq:pgm-optimality-1}\\
&\qquad {} = \frac{1}{d} \sum_{\mu = \beta+\square} x_{\mu,\beta}^{-1}  \tr\left[ P_{\mu} \left( \one_{A_1} \otimes |\beta,\tilde{q}_\beta,\tilde{p}_\beta\rangle \langle \alpha,q_\alpha,p_\alpha|_{A_1^c}\right)\right] \label{eq:pgm-optimality-15}\\
&\qquad {} = \frac{1}{d} \sum_{\mu = \beta+\square} x_{\mu,\beta}^{-1}  \tr\left[ \tr_{A_1} (P_{\mu}) |\beta,\tilde{q}_\beta,\tilde{p}_\beta\rangle \langle \alpha,q_\alpha,p_\alpha|_{A_1^c} \right]\\
&\qquad {} = d^{N-1}  \left( \sum_{\mu'=\beta+\square} \sqrt{m_{\mu'} d_{\mu'}}\right)^{-1} m_\beta \sum_{\mu=\beta + \square} \frac{\sqrt{d_{\mu}}}{\sqrt{m_{\mu}}} \frac{m_{\mu}}{m_\beta} \delta_{\alpha,\beta} \delta_{q_\alpha,\tilde{q}_\alpha} \delta_{p_\alpha,\tilde{p}_\alpha}  \label{eq:pgm-optimality-2}\\
&\qquad {} = d^{N-1} \delta_{\alpha,\beta}  \delta_{q_\alpha,\tilde{q}_\alpha} \delta_{p_\alpha,\tilde{p}_\alpha},
\end{align}
where step \eqref{eq:pgm-optimality-1} uses \eqref{eq:U-action-Phi+}, 
step \eqref{eq:pgm-optimality-15} uses the identity $P'_{\alpha'}|\beta,\tilde{q}_\beta,\tilde{p}_\beta\rangle_{A_1^c} = \delta_{\alpha',\beta}|\beta,\tilde{q}_\beta,\tilde{p}_\beta\rangle_{A_1^c}$ and a partial trace over $B$, 
and step \eqref{eq:pgm-optimality-2} uses \Cref{lem:Young-proj-partial-trace}, the formula \eqref{eq:x-coefficients-formula} for $x_{\mu,\beta}$, and another application of $P'_{\hat{\alpha}}|\beta,\tilde{q}_\beta,\tilde{p}_\beta\rangle_{A_1^c} = \delta_{\hat{\alpha},\beta}|\beta,\tilde{q}_\beta,\tilde{p}_\beta\rangle_{A_1^c}$.

The above calculation proves \eqref{eq:inverse-innerproduct}, and thus $X \geq \rho_1$ follows from \Cref{lem:support-vector-psd}.
Hence, $X$ is feasible in the dual program \eqref{eq:state-discrimination-dual}, which concludes the proof of optimality of the pretty good measurement for the PBT protocol with $N$ maximally entangled states.

\section{Fully optimized protocol}\label{sec:optimized}

We now turn our attention to the fully optimized PBT protocol.
In this case, we seek to find a port state $\phi_{A^NB^N}$ and POVM $E=\lbrace E_i\rbrace_{i=1}^N$ such that the entanglement fidelity $F(\Lambda)$ for the corresponding teleportation channel $\Lambda$ defined in \eqref{eq:teleportation-channel} is maximized.

The port state $\phi_{A^NB^N}$ in a PBT protocol can always be assumed to be pure \cite{majenz2018phd,christandl2021asymptotic}.
Fixing the marginal $\phi_{A^N}$ on $A^N$, we further assume without loss of generality that $|\phi\rangle_{A^NB^N}$ is the ``canonical'' purification of $\phi_{A^N}$,\footnote{Any two purifications of $\phi_{A^N}$ on $A^NB^N$ are related by an isometry acting on $B^N$, so we may assume that Bob applies a suitable isometry on $B^N$ to obtain the state in \eqref{eq:steering} before starting the protocol.
	The entanglement fidelity of the resulting protocol will be no worse than the original one.}
\begin{align}
|\phi\rangle_{A^NB^N} = (O_{A^N} \otimes \one_{B^N}) |\Phi^+\rangle_{AB}^{\otimes N},\label{eq:steering}
\end{align}
where the positive semidefinite operator $O_{A^N} = \sqrt{d^N \phi_{A^N}}$ satisfies $\tr O_{A^N}^\dagger O_{A^N} = \tr O_{A^N}^2 = d^N$. 
According to \Cref{sec:pbt}, PBT using the state $\phi_{A^NB^N}$ is equivalent to discriminating the states
\begin{align}
\eta_i = \tr_{B_i^c} \phi_{A^NB^N} = O_{A^N} \left( \Phi^+_{A_iB_i} \otimes \pi_{A_1^c} \right) O_{A^N}^\dagger,\label{eq:eta-i}
\end{align}
each drawn uniformly at random with probability $\frac{1}{N}$.

We saw in \Cref{sec:standard} that the states $\rho_i = \tr_{B_i^c} \left(\Phi^+_{AB}\right)^{\otimes N}$ have $U^{\otimes N}\otimes \bar{U}$ and $S_{N-1}$ symmetries, which facilitated the calculation of the entanglement fidelity of the corresponding PBT protocol.
\textcite{majenz2018phd} showed that these symmetries can \emph{always} be assumed in an arbitrary PBT protocol (see also the extended discussion in \cite[Sec.~3.3]{christandl2021asymptotic}).
More precisely, we may assume without loss of generality that $\phi_{A^N}$ (or equivalently, $\phi_{B^N}$) is a symmetric Werner state, which implies the following symmetries for $O_{A^N} = \sqrt{d^N\phi_{A^N}}$:
\begin{align}
\left[U^{\otimes N},O_{A^N}\right] &= 0 \quad\text{for all $U\in\cU_d$,} \label{eq:O-U-symmetry}\\
\left[\pi,O_{A^N}\right] &= 0 \quad\text{for all $\pi\in S_N$.}\label{eq:O-S_N-symmetry}
\end{align}
We conclude that, similar to \Cref{sec:standard}, the states $\eta_i$ on $A^NB$ defined in \eqref{eq:eta-i} satisfy
\begin{align}
\left[U^{\otimes N}\otimes \bar{U},\eta_i\right] &= 0 \quad\text{for all $U\in\cU_d$,} \label{eq:eta-U-symmetry}\\
\left[\varphi\otimes \one_{A_iB},\eta_i \right] &= 0 \quad\text{for all $\varphi\in S_{N-1}$,}\label{eq:eta-S_N-1-symmetry}\\
\pi \eta_i \pi^\dagger &= \eta_{\pi(i)} \quad\text{for all $\pi\in S_N$,}\label{eq:eta-orbit}
\end{align}
where in \eqref{eq:eta-S_N-1-symmetry} the action of $S_{N-1}$ is defined on $A_i^c$.

\textcite{mozrzymas2018optimal} showed that the entanglement fidelity of the fully optimized PBT protocol is given by the expression
\begin{align}
F(\Lambda) = \frac{1}{d^{N+2}} \max_{\lbrace c_\mu\rbrace} \sum_{\alpha\vdash_d N-1} \left(\sum_{\mu=\alpha + \square} \sqrt{c_\mu d_\mu m_{d,\mu}}\right)^2,
\label{eq:mozrzymas}
\end{align}
where the non-negative coefficients $\lbrace c_\mu\rbrace_{\mu\vdash_d N}$ satisfy 
\begin{align}
\sum_{\mu\vdash_d N} c_\mu d_\mu m_{d,\mu} = d^N. \label{eq:cmu-condition}
\end{align}
We will rederive \eqref{eq:mozrzymas} in this section.
Somewhat surprisingly, it will turn out that the \emph{same} pretty good measurement as used in \Cref{sec:standard} maximizes the success probability of distinguishing the states $\eta_i$, and hence also achieves the optimal value \eqref{eq:mozrzymas} for the entanglement fidelity via \eqref{eq:PBT-state-discrimination}.

\subsection{Performance of the pretty good measurement}

We consider again the pretty good measurement $E=\lbrace E_i\rbrace_{i=1}^N$ with $E_i = \brho^{-1/2}\rho_i\brho^{-1/2}$, defined in terms of the states $\rho_i$ given in \eqref{eq:rho_i}.
These states differ from the $\eta_i$ in \eqref{eq:eta-i} above by the conjugation with the operator $O_{A^N}$.
We stress that $E$ is \emph{not} the pretty good measurement defined in terms of the states $\eta_i$, which would be a sub-optimal choice.

The success probability of distinguishing the state ensemble $\lbrace (\frac{1}{N},\eta_i)\rbrace_{i=1}^N$ with the pretty good measurement $E=\lbrace E_i\rbrace_{i=1}^N$ is equal to
\begin{align}
\psucc &= \frac{1}{N} \sum_{i=1}^N \tr \left(\eta_i E_i \right)
= \frac{1}{N} \sum_{i=1}^N \tr\left(O_{A^N} \rho_i O_{A^N}^\dagger \brho^{-1/2} \rho_i \brho^{-1/2}\right),
\end{align}
where we inserted $\eta_i = O_{A^N} \rho_i O_{A^N}^\dagger$.
In analogy to \Cref{sec:standard}, we define an operator 
\begin{align}
Y = \sum_{i=1}^N O_{A^N} \rho_i O_{A^N}^\dagger \brho^{-1/2} \rho_i \brho^{-1/2},\label{eq:Y}
\end{align}
satisfying $\frac{1}{N}\tr Y = \psucc$.
Due to the symmetries of the states $\rho_i$ (\cref{eq:rho_i-symmetry-U,eq:rho_i-symmetry-SN-1,eq:rho_i-orbit}), the state $\brho$~(\cref{eq:brho-symmetry-U,eq:brho-symmetry-SN}), the operator $O_{A^N}$ (\cref{eq:O-U-symmetry,eq:O-S_N-symmetry}), and the states $\eta_i$ (\cref{eq:eta-U-symmetry,eq:eta-S_N-1-symmetry,eq:eta-orbit}), we infer that $Y$ has the following symmetries:
\begin{align}
\left[U^{\otimes N} \otimes \bar{U},Y \right] &= 0 \quad\text{for all $U\in \cU_d$,} \label{eq:Y-symmetry-U}\\
\left[\pi\otimes \one_B,Y\right] &= 0 \quad\text{for all $\pi\in S_{N}$,} \label{eq:Y-symmetry-SN}
\end{align}
such that we can again write $Y$ in the form\footnote{In analogy to the discussion about the operator $X$ in \Cref{sec:standard}, one can show that the coefficients $y_{\mu,i}$ defined with respect to the decomposition \eqref{eq:Cd-N+1-decomposition} vanish whenever $\mu-\eps_i$ does \emph{not} correspond to a Young diagram $\alpha\vdash_d N-1$.}
\begin{align}
Y = \bigoplus_{\alpha\vdash_d N-1}\, \bigoplus_{\mu = \alpha + \square} y_{\mu,\alpha}\, \one_{V_\alpha^d} \otimes \one_{W_\mu}.
\label{eq:Y-decomposition}
\end{align}

In the following we determine the value of $\tr Y$ and a formula for the coefficients $y_{\mu,\alpha}$ appearing in \eqref{eq:Y-decomposition}.
We again abbreviate $m_\mu\equiv m_{d,\mu}$ for the dimension of the Weyl module $V_\mu^d$.
As before, we denote by $P_\mu$ for $\mu\vdash_d N$ the projection onto the summand $V_\mu^d\otimes W_\mu$ in the Schur-Weyl decomposition \eqref{eq:schur-weyl-duality}, by $Q_\alpha$ for $\alpha\vdash_d N-1$ the isotypical projection for the $\cU_d$ action by $U^{\otimes N}\otimes \bar{U}$ as defined via decomposition \eqref{eq:Cd-N+1-decomposition}, and by $P'_\alpha$ for $\alpha\vdash_d N-1$ the isotypical projection with respect to the action of $\cU_d$ on $A_1^c$ by $U^{\otimes N-1}$.

We first use the $U^{\otimes N}$ and $S_N$ symmetries of $O_{A^N}$ (\cref{eq:O-U-symmetry,eq:O-S_N-symmetry}) to write it as
\begin{align}
O_{A^N} = \bigoplus_{\mu\vdash_d N} \sqrt{c_\mu}\, \one_{V_\mu^d}\otimes \one_{W_\mu},
\label{eq:O-decomposition}
\end{align}
where $\lbrace c_\mu\rbrace_{\mu\vdash_d N}$ are non-negative coefficients (recall that $O_{A^N}=\sqrt{d^N \phi_{A^N}}$ is positive semidefinite).
Since $\tr O_{A^N}^2 = d^N$, we have $\sum_{\mu\vdash_d N} c_\mu d_\mu m_\mu = d^N$, which is precisely the condition \eqref{eq:cmu-condition} for the coefficients $c_{\mu}$ in the expression \eqref{eq:mozrzymas} for the entanglement fidelity.

We are now ready to compute the trace of $Y$.
By symmetry, and using the expressions for $\brho$ from \Cref{lem:eigenvalues-average-state} and for $O_{A^N}$ in \eqref{eq:O-decomposition}, we have
\begin{align}
&\tr \left[Y \left( P_\mu\otimes \one_B \right) Q_\alpha\right] \notag\\
&\qquad {} = N \tr\left[ O_{A^N} \rho_1 O_{A^N}^\dagger \brho^{-1/2} \rho_1 \brho^{-1/2} \left(P_\mu \otimes \one_B\right) Q_\alpha \right]\\
&\qquad {} = \frac{N}{d^{2N-2}} \sum_{\alpha',\alpha''\vdash_d N-1} \, \sum_{\substack{\mu'=\alpha'+\square\\ \mu''=\alpha''+\square}} \, \sum_{\lambda',\lambda''\vdash_d N} r_{\mu',\alpha'}^{-1/2} r_{\mu'',\alpha''}^{-1/2}\, \sqrt{c_{\lambda'} c_{\lambda''}}\notag\\
&\qquad \phantom{=} {} \times \tr\left[(P_{\lambda'} \otimes \one_B)\, \Phi^+_{A_1B} (P_{\lambda''} \otimes \one_B) (P_{\mu'} \otimes \one_B) Q_{\alpha'}\,\Phi^+_{A_1B} (P_{\mu''} \otimes \one_B) Q_{\alpha''} (P_{\mu} \otimes \one_B) Q_{\alpha}\right] \label{eq:y-coeff-inter1}\\
&\qquad {} = \frac{N}{d^{2N-2}} \, r_{\mu,\alpha}^{-1/2} \sqrt{c_\mu}\sum_{\mu'=\alpha+\square} r_{\mu',\alpha}^{-1/2} \, \sqrt{c_{\mu'}} \tr\left[\Phi^+_{A_1B} (P_{\mu'} \otimes \one_B) \Phi^+_{A_1B} (P_{\mu} \otimes \one_B) (\one_{A_1B} \otimes P'_{\alpha})\right] \label{eq:y-coeff-inter2}\\
&\qquad {} = \frac{N}{d^{2N}} r_{\mu,\alpha}^{-1/2} \sqrt{c_\mu}\sum_{\mu'=\alpha+\square} r_{\mu',\alpha}^{-1/2} \, \sqrt{c_{\mu'}}\, \tr\left[ \tr_{A_1} (P_{\mu'}) \tr_{A_1} (P_{\mu}) P'_{\alpha}\right] \label{eq:y-coeff-inter3}\\
&\qquad {} = \frac{1}{d^{N}} \frac{ \sqrt{c_\mu m_\alpha d_\mu } }{\sqrt{m_\mu d_\alpha}} \sum_{\mu'=\alpha+\square} \frac{ \sqrt{c_{\mu'} m_\alpha d_{\mu'} } }{\sqrt{m_{\mu'} d_\alpha}} \frac{m_\mu m_{\mu'}}{m_\alpha^2} d_\alpha m_\alpha \label{eq:y-coeff-inter4} \\
&\qquad {} = \frac{1}{d^{N}} \sqrt{c_\mu m_\mu d_\mu }  \sum_{\mu'=\alpha+\square} \sqrt{c_{\mu'} m_{\mu'} d_{\mu'} } \label{eq:y-coeff-inter5}.
\end{align}
In step \eqref{eq:y-coeff-inter2} we used \eqref{eq:U-action-Phi+} for the terms $\Phi^+_{A_1B}Q_*$ and orthogonality among the projectors $P_*$ and $P'_*$, respectively, and in step \eqref{eq:y-coeff-inter4} we again used \Cref{lem:Young-proj-partial-trace} in the same way as in \Cref{sec:standard}.

The trace of $Y$ is obtained by summing \eqref{eq:y-coeff-inter5} over $\alpha\vdash_d N-1$ and $\mu = \alpha + \square$, giving
\begin{align}
\tr Y &= \sum_{\alpha\vdash_d N-1} \sum_{\mu=\alpha + \square} \tr \left[Y \left( P_\mu\otimes \one_B \right) Q_\alpha\right]\\
&= \frac{1}{d^N} \sum_{\alpha\vdash_d N-1} \left(\sum_{\mu=\alpha+\square} \sqrt{c_{\mu} m_{\mu} d_{\mu} }\right)^2.\label{eq:opt-psucc}
\end{align}
Maximizing \eqref{eq:opt-psucc} over all non-negative coefficients $\lbrace c_\mu\rbrace_{\mu\vdash_d N}$ satisfying $\sum_{\mu\vdash_d N} c_\mu m_\mu d_\mu = d^N$ and using \eqref{eq:PBT-state-discrimination} together with $\psucc = \frac{1}{N} \tr Y$ now proves that the entanglement fidelity of the PBT protocol $(\phi_{A^NB^N},E)$ is given by formula \eqref{eq:mozrzymas} derived in \cite{mozrzymas2018optimal}.
Here, $\phi_{A^NB^N}$ is defined via \eqref{eq:steering} and \eqref{eq:O-decomposition}, and the pretty good measurement $E=\lbrace E_i\rbrace_{i=1}^N$ is defined in terms of the states $\rho_i$ as given in \eqref{eq:rho_i}.

It remains to determine the coefficients $y_{\mu,\alpha}$ appearing in \eqref{eq:Y-decomposition}.
By definition, for Young diagrams $\alpha\vdash_d N-1$ and $\mu = \alpha + \square$,
\begin{align}
\tr \left[Y \left( P_\mu\otimes \one_B \right) Q_\alpha\right] = y_{\mu,\alpha} m_\alpha d_\mu.
\end{align}
This is equal to \eqref{eq:y-coeff-inter5} by the above calculation, leading to the following formula for the $y_{\mu,\alpha}$:
\begin{align}
y_{\mu,\alpha} = \frac{1}{d^{N}} \frac{1}{m_\alpha d_\mu} \sqrt{c_\mu m_\mu d_\mu}  \sum_{\mu'=\alpha+\square} \sqrt{c_{\mu'} m_{\mu'} d_{\mu'} }
\label{eq:y-coefficients-formula}
\end{align}

We stress that in expression \eqref{eq:opt-psucc} the port state $\phi_{A^NB^N}$ is optimized over via the coefficients $\lbrace c_\mu\rbrace_{\mu\vdash_d N}$, while the POVM is \emph{fixed} to be the pretty good measurement $E$ discriminating the states $\rho_i$ in \eqref{eq:rho_i}.
We show in the next section that this measurement $E$ is in fact optimal for any given port state $\phi_{A^NB^N}$ defined via \eqref{eq:steering}, \eqref{eq:O-decomposition}, and the coefficients $\lbrace c_\mu\rbrace_{\mu\vdash_d N}$, which also proves optimality of $E$ for the optimal such $\phi_{A^NB^N}$.

\subsection{Optimality of the pretty good measurement}

It remains to show that the choice of the pretty good measurement $E$ associated with $\lbrace (\frac{1}{N},\rho_i)\rbrace_{i=1}^N$ achieves the optimal success probability of discriminating the state ensemble $\lbrace (\frac{1}{N},\eta_i)\rbrace_{i=1}^N$.
To prove this, we follow a similar strategy as in \Cref{sec:standard}: 
Once we establish that the operator $\frac{1}{N}Y$ with $Y$ as defined in \eqref{eq:Y} is feasible for the dual program \eqref{eq:state-discrimination-dual}, optimality follows immediately from weak duality.

Feasibility of $\frac{1}{N}Y$ is equivalent to $Y \geq \eta_i$ for all $i=1,\dots, N$, where $\eta_i = O_{A^N} \rho_i O_{A^N}$ with 
\begin{align} 
	O_{A^N} = \bigoplus_{\mu\vdash_d N} \sqrt{c_\mu}\, \one_{V_\mu^d}\otimes \one_{W_\mu}.\label{eq:O}
\end{align}
By symmetry, it suffices to show that $Y \geq \eta_1$, for which we once more make use of \Cref{lem:support-vector-psd}.
First, we recall the expression \eqref{eq:Y-decomposition} for the operator $Y$, which together with formula \eqref{eq:y-coefficients-formula} for the coefficients $y_{\mu,\alpha}$ shows that $Y$ is positive semidefinite (recall that $c_\mu\geq 0$ for all $\mu\vdash_d N$).
As the collection of vectors in \Cref{lem:support-vector-psd}, we choose
\begin{align}
	|\chi(\alpha,q_\alpha,p_\alpha)\rangle \coloneqq (O_{A^N} \otimes \one_B) \left(|\Phi^+\rangle_{A_1B} \otimes |\alpha,q_\alpha,p_\alpha\rangle_{A_1^c}\right)\label{eq:chi-vectors}
\end{align}
for $\alpha\vdash_d N-1$, $1\leq q_\alpha\leq m_\alpha$, and $1\leq p_\alpha\leq d_\alpha$, where $|\alpha,q_\alpha,p_\alpha\rangle_{A_1^c}$ is the Schur basis on $A_1^c$ (see \Cref{sec:standard-optimality}) and $O_{A^N}$ is the operator in \eqref{eq:O}.
Because of the spectral decomposition \eqref{eq:rho-in-schur-basis} of $\rho_1$ and \eqref{eq:chi-vectors}, we have
\begin{align}
	\eta_1 =  O_{A^N} \rho_1 O_{A^N} = \frac{1}{d^{N-1}} \sum_{\alpha\vdash_d N-1} \sum_{q_\alpha,p_\alpha} |\chi(\alpha,q_\alpha,p_\alpha)\rangle \langle \chi(\alpha,q_\alpha,p_\alpha)|,
\end{align}
so that $Y \geq \eta_1$ will follow from \Cref{lem:support-vector-psd} once we establish that
\begin{align}
	&\langle \chi(\alpha,q_\alpha,p_\alpha)| Y^{-1}| \chi(\beta,\tilde{q}_\beta,\tilde{p}_\beta)\rangle\notag\\ 
	& \qquad {} = 
	\left(\langle\Phi^+|_{A_1B} \otimes \langle\alpha,q_\alpha,p_\alpha|_{A_1^c}\right) O_{A^N}  Y^{-1} O_{A^N} \left(|\Phi^+\rangle_{A_1B} \otimes |\beta,\tilde{q}_\beta,\tilde{p}_\beta\rangle_{A_1^c}\right)
	\\ & \qquad {} =\delta_{\alpha,\beta} \delta_{q_\alpha,\tilde{q}_\alpha} \delta_{p_\alpha,\tilde{p}_\alpha} d^{N-1}\label{eq:Y-inner-product}
\end{align}
holds for all $\alpha,\beta\vdash_d N-1$, $1\leq q_\alpha\leq m_\alpha$, $1\leq \tilde{q}_\beta \leq m_\beta$, $1\leq p_\alpha\leq d_\alpha$, and $1\leq \tilde{p}_\beta\leq d_\beta$.

To this end, we first observe that
\begin{align}
 O_{A^N} Y^{-1} O_{A^N} = \bigoplus_{\alpha\vdash_d N-1}\, \bigoplus_{\mu=\alpha + \square} c_\mu y_{\mu,\alpha}^{-1}\,\one_{V_\alpha^d}\otimes \one_{W_\mu}.
\end{align}
We then compute:
\begin{align}
&\langle \chi(\alpha,q_\alpha,p_\alpha)| Y^{-1} | \chi(\beta,\tilde{q}_\beta,\tilde{p}_\beta)\rangle \notag\\
&\qqquad {} = \sum_{\alpha'\vdash_d N-1}\, \sum_{\mu = \alpha'+\square} c_{\mu} y_{\mu,\alpha'}^{-1} \tr\left[ (P_{\mu}\otimes \one_B) Q_{\alpha'} \left( \Phi^+_{A_1B}  \otimes |\beta,\tilde{q}_\beta,\tilde{p}_\beta\rangle \langle \alpha,q_\alpha,p_\alpha|_{A_1^c}\right) \right]\\
&\qqquad {} = \frac{1}{d} \sum_{\mu=\beta + \square} c_{\mu} y_{\mu,\beta}^{-1} \tr\left[ \tr_{A_1} (P_{\mu}) |\beta,\tilde{q}_\beta,\tilde{p}_\beta\rangle \langle \alpha,q_\alpha,p_\alpha|_{A_1^c}\right]\\
&\qqquad {} = d^{N-1} \left(\sum_{\mu'=\beta+\square} \sqrt{c_{\mu'} m_{\mu'} d_{\mu'}}\right)^{-1} \sum_{\mu = \beta + \square} \frac{c_{\mu} m_\beta d_{\mu}}{\sqrt{c_{\mu} m_{\mu} d_{\mu}}} \frac{m_{\mu}}{m_\beta} \delta_{\alpha,\beta} \delta_{q_\alpha,\tilde{q}_\alpha} \delta_{p_\alpha,\tilde{p}_\alpha}\label{eq:pgm-optimality-3}\\
&\qqquad {} = d^{N-1} \delta_{\alpha,\beta} \delta_{q_\alpha,\tilde{q}_\alpha} \delta_{p_\alpha,\tilde{p}_\alpha},
\end{align}
where we used similar arguments as in \Cref{sec:standard-optimality}, and the expression \eqref{eq:y-coefficients-formula} for the coefficients $y_{\mu,\beta}$ in step \eqref{eq:pgm-optimality-3}.

This proves \eqref{eq:Y-inner-product}, so that $Y \geq \eta_1$ follows from \Cref{lem:support-vector-psd}.
Hence, $\frac{1}{N}Y$ is feasible in the dual program \eqref{eq:state-discrimination-dual}, which proves that the pretty good measurement from \Cref{sec:standard} optimally distinguishes the states $\eta_i$ defined in \eqref{eq:eta-i} in terms of an arbitrary set of non-negative coefficients $\lbrace c_\mu\rbrace_{\mu\vdash_d N}$ satisfying $\sum_{\mu\vdash_d N} c_\mu m_\mu d_\mu = d^N$.
We showed above that the optimal success probability of this discrimination problem as a function of $\lbrace c_\mu\rbrace_{\mu\vdash_d N}$ is equal to \eqref{eq:opt-psucc}.
Optimizing over the coefficients $\lbrace c_\mu\rbrace_{\mu\vdash_d N}$ and using \eqref{eq:PBT-state-discrimination} then leads to the expression \eqref{eq:mozrzymas} for the entanglement fidelity of the fully optimized PBT protocol.

\section{Discussion}\label{sec:discussion}

In this paper we proved that the pretty good measurement is optimal for PBT protocols using maximally entangled states.
Furthermore, we showed that the very same measurement also achieves the optimal entanglement fidelity for arbitrary port states once the natural symmetries of PBT have been imposed without loss of generality.
We stress once again that for the second result the pretty good measurement is \emph{not} derived from the optimal port state (see \Cref{sec:pbt} for how to obtain the state discrimination problem from a given port state), but instead from $N$ maximally entangled states.

In the course of proving optimality of the pretty good measurement, we also rederived the representation-theoretic formulas for the entanglement fidelity of PBT protocols using maximally entangled states \cite{studzinski2017port} and using an optimized port state \cite{mozrzymas2018optimal}.
In order to better distinguish the two settings, we adopt the notation of \cite{christandl2021asymptotic} and write $\Fstd_d(N)$ and $F_d^*(N)$ for the entanglement fidelity in each case, respectively.\footnote{In \cite{christandl2021asymptotic}, the PBT protocol based on $N$ maximally entangled states and the associated pretty good measurement is called the \emph{standard protocol}.}
This notation makes the dependence of $F$ on the local dimension $d$ and the number of ports $N$ explicit, and it highlights the assumption of fixed but arbitrary $d$ and varying $N$ made in this paper as well as in \cite{christandl2021asymptotic}.

In \Cref{sec:standard} we rederived the following result from \cite{studzinski2017port}:
\begin{align}
\Fstd_d(N) = \frac{1}{d^{N+2}} \sum_{\alpha\vdash_d N-1} \left(\sum_{\mu=\alpha + \square} \sqrt{d_{\mu} m_{\mu}}\right)^2.
\label{eq:Fstd}
\end{align}
\textcite{ishizaka2008asymptotic} (see also \cite{beigi2011simplified}) proved that $\Fstd_d(N) \geq 1-\frac{d^2-1}{N}$, which shows that the PBT protocol becomes perfect in the limit $N\to\infty$ for fixed $d$.
One of the main goals of \cite{christandl2021asymptotic} was to determine the exact first-order coefficient of this convergence.
We showed in \cite{christandl2021asymptotic} that, for any $\delta>0$,
\begin{align}
\Fstd_d(N) = 1 - \frac{d^2-1}{4N} + O(N^{-3/2+\delta}).\label{eq:Fstd-asymptotics}
\end{align}

For optimal PBT, we rederived in \Cref{sec:optimized} the following expression for the entanglement fidelity first proved in \cite{mozrzymas2018optimal}:
\begin{align}
F^*_d(N) = \frac{1}{d^{N+2}} \max_{\lbrace c_\mu\rbrace} \sum_{\alpha\vdash_d N-1} \left(\sum_{\mu=\alpha+\square} \sqrt{c_{\mu}m_{\mu} d_{\mu}}\right)^2,
\label{eq:F*}
\end{align}
where the non-negative coefficients $\lbrace c_\mu\rbrace_{\mu\vdash_d N}$ satisfy $\sum_{\mu\vdash_d N} c_\mu m_\mu d_\mu = d^N$.
\Cref{eq:F*} bears a striking resemblance with \eqref{eq:Fstd}, and the additional optimization over coefficients $\lbrace c_\mu\rbrace_{\mu\vdash_d N}$ corresponds to the optimization over the port state (see \Cref{sec:optimized} for details).
\textcite{ishizaka2015some} proved the upper bound $F^*_d(N) \leq 1- c_d N^{-2} + O(N^{-3})$ with $c_d = (4(d-1))^{-1}$, which was improved by \textcite{majenz2018phd} (see also \cite{christandl2021asymptotic}) to $c_d = (d^2-1)/8$ whenever $N >\frac{d^2}{2}$.
However, prior to our work \cite{christandl2021asymptotic} it was not clear whether there are protocols achieving the $N^{-2}$ scaling asymptotically.
In \cite{christandl2021asymptotic}, we exhibited a protocol with such a scaling in $N$, albeit with non-matching coefficients in $d$.
This resulted in the asymptotic expansion
\begin{align}
F^*_d(N) = 1 - \Theta(N^{-2}),\label{eq:F*-asymptotics}
\end{align}
where $f=\Theta(g)$ means that both $f=O(g)$ and $g=O(f)$.
It remains a challenging open problem to determine the exact coefficient of $N^{-2}$ as a function of $d$ in \eqref{eq:F*-asymptotics}.
Moreover, an investigation of \eqref{eq:F*} in the interesting limit $N,d\to\infty$ with $N/d^2$ fixed has yet to be carried out.

In this paper we only discussed the ``deterministic'' variant of PBT, in which the protocol gives an output state that approximates the target state.
In ``probabilistic'' PBT the protocol yields an exact copy of the target state, but only succeeds with a certain success probability \cite{ishizaka2008asymptotic,ishizaka2009quantum}.
A description of this probability in terms of representation-theoretic data was obtained in \cite{studzinski2017port,mozrzymas2018optimal}, along with converse bounds \cite{pitalua2013deduction} and asymptotic expansions \cite{mozrzymas2018optimal,christandl2021asymptotic}.
It should be a stimulating exercise to apply the techniques of \cite{christandl2021asymptotic} and the present paper to rederive the results on probabilistic PBT proved in \cite{studzinski2017port,mozrzymas2018optimal}.
Moreover, a ``multi-port'' generalization of PBT was recently proposed in \cite{studzinski2020efficient,kopszak2021multiport,mozrzymas2021optimal}, and the methods employed here could potentially be applied to study this generalized setting as well.
Finally, it would be interesting to derive expressions for the optimal entanglement fidelity in the case of \emph{noisy} maximally entangled states, e.g., when each maximally entangled state is shared between Alice and Bob via a noisy quantum channel.

\paragraph*{Acknowledgments.}
I would like to thank Christian Majenz, Connor Paul-Paddock, and Michael Walter for valuable discussions and helpful feedback. 
I am also grateful to the anonymous referee for useful comments on an earlier version of this manuscript, and permission to reproduce their proof of \Cref{lem:support-vector-psd}.
This research was partially funded through the Army Research Lab CDQI program.

\paragraph*{Conflict of interest.} The corresponding author states that there is no conflict of interest.

\appendix
\section{\texorpdfstring{Proof of \Cref{lem:support-vector-psd}}{Proof of Lemma 1}}\label{sec:app-lemma}
In this appendix we give a proof of \Cref{lem:support-vector-psd}, which is restated below for convenience.
For a positive semidefinite operator $X$, the generalized inverse $X^{-1}$ and the orthogonal projection $\Pi_X$ onto $\supp X$ are defined as in \Cref{sec:preliminaries}.
Note that $\supp X = (\ker X)^\perp = \im X$ for positive semidefinite $X$.

\begin{support-vector-psd}[restated]
	\restateSupportVectorPSD
\end{support-vector-psd}

\begin{proof}
We prove this lemma by induction on $K$.

Let first $K=1$.
Recall that $\Pi_X |\xi\rangle = |\xi\rangle$ by assumption, and $\Pi_X = \sqrt{X}\sqrt{X^{-1}}$ by the definitions of the square root and generalized inverse of $X$.
For any $|\psi\rangle\in\cH$,
\begin{align}
	\langle \psi| \xi\rangle\langle\xi|\psi\rangle = |\langle \psi|\xi\rangle|^2 = |\langle \psi| \Pi_X|\xi\rangle|^2 &= |\langle \psi|\sqrt{X}\sqrt{X^{-1}}|\xi\rangle|^2\\
	&\leq \langle \psi | X |\psi\rangle \langle \xi | X^{-1} | \xi\rangle\label{eq:inequality}\\
	& = c \langle \psi | X |\psi\rangle,
\end{align}
where \eqref{eq:inequality} follows from the Cauchy-Schwarz inequality.
Then $|\xi\rangle\langle \xi| \leq cX$ holds since $|\psi\rangle\in\cH$ was arbitrary.
Taking traces on both sides of this operator inequality and using $|\xi\rangle \neq 0$ and $X\geq 0$ shows $c>0$, from which the induction base case $\frac{1}{c}|\xi\rangle\langle \xi| \leq X$ follows.

Let now $K>1$.
Applying the argument above to $|\xi_K\rangle$ shows that $Y\coloneqq X - c^{-1}|\xi_K\rangle\langle\xi_K|$ is positive semidefinite.
In order to use the induction hypothesis, we need to verify that (a) $|\xi_j\rangle\in \im(Y)$ for $1\leq j\leq K-1$ and (b) $\langle \xi_j|Y^{-1}|\xi_k\rangle = \delta_{j,k} c$ for $1\leq j,k\leq K-1$.

To show (a), observe that for any $1\leq j\leq K-1$,
\begin{align}
	YX^{-1} |\xi_j\rangle = \left(X-\frac{1}{c}|\xi_K\rangle\langle \xi_K|\right)X^{-1} |\xi_j\rangle = XX^{-1} |\xi_j\rangle - \frac{1}{c} |\xi_K\rangle \langle \xi_K|X^{-1}|\xi_j\rangle = \Pi_X |\xi_j\rangle = |\xi_j\rangle,\label{eq:image}
\end{align}
since $\langle \xi_K|X^{-1}|\xi_j\rangle = 0$ and $|\xi_j\rangle\in\im(X)$ for $1\leq j\leq K-1$ by assumption.

To show (b), we apply $Y^{-1}$ to both sides of \eqref{eq:image}, giving
\begin{align}
	Y^{-1}|\xi_j\rangle = Y^{-1}YX^{-1}|\xi_j\rangle = \Pi_YX^{-1} |\xi_j\rangle.
\end{align}
Taking the inner product with any $|\xi_k\rangle$ for $1\leq k\leq K-1$ and using (a) then shows that
\begin{align}
	\langle \xi_k | Y^{-1} |\xi_j\rangle = \langle \xi_k| \Pi_Y X^{-1} |\xi_j\rangle = \langle \xi_k | X^{-1} |\xi_j\rangle = \delta_{j,k} c.
\end{align}

We may therefore apply the induction hypothesis to $Y$ and the vectors $\lbrace |\xi_j\rangle\rbrace_{j=1}^{K-1}$, giving
\begin{align}
	Y = X - \frac{1}{c} |\xi_K\rangle\langle \xi_K| \geq \frac{1}{c} \sum_{j=1}^{K-1} |\xi_j\rangle\langle \xi_j|.
\end{align}
Rearranging this inequality yields the assertion of \Cref{lem:support-vector-psd}.
\end{proof}

\emergencystretch=1em
\renewcommand*{\bibfont}{\small}
\printbibliography[heading=bibintoc]
\end{document}